\documentclass[12pt,preprint]{aastex}
\usepackage{epsfig}
                                                                                
\def\vkm{km s$^{-1}$}

\def\degree{$^\circ$}
\def\arcs#1{$#1''$}
\def\arcsa#1#2{$#1^{\prime\prime}_{^\textrm{.}}#2$}

\def\solarmass{$M_\odot$}

\def\Jyb{Jy beam$^{-1}$}
\def\mJyb{mJy beam$^{-1}$}
\def\Jybk{Jy beam$^{-1}$ km s$^{-1}$}
\def\mJybk{mJy beam$^{-1}$ km s$^{-1}$}

\def\cmc{cm$^{-3}$}
\def\cms{cm$^{-2}$}
\def\micron{$\mu$m}
\def\VLSR{V_\textrm{\scriptsize LSR}}
\def\Vsys{V_\textrm{\scriptsize sys}}
\def\Voff{V_\textrm{\scriptsize off}}

\def\mHt{m_{\textrm{\scriptsize H}_2}}

\def\NHt{N_{\textrm{\scriptsize H}_2}}

\def\H2{H$_2$}
\def\N2HP{N$_2$H$^+$}
\def\HCOP{HCO$^+$}

\def\NH3{NH$_3$}

\def\SOTwo{SO$_2$}

\def\CHtDOH{CH$_2$DOH}

\def\SOt{$N_J=8_9-7_8$}

\def\SOO{SO$_2$}
\def\HCOP{HCO$^+$}

\def\putfig#1#2#3{\epsfig{scale=#1,angle=#2,figure=#3}}
\def\putfiga#1#2#3{}
\def\leftblank#1{}

\begin{document}

\title{A 100 au-Wide Bipolar Rotating Shell Emanating From The HH 212
Protostellar Disk: A Disk Wind?}


\author{Chin-Fei Lee\altaffilmark{1,2}, Zhi-Yun Li\altaffilmark{3}, Claudio
Codella\altaffilmark{4}, Paul T.P.  Ho\altaffilmark{1}, Linda
Podio\altaffilmark{4}, Naomi Hirano\altaffilmark{1}, Hsien
Shang\altaffilmark{1}, Neal J. Turner\altaffilmark{5}, and Qizhou
Zhang\altaffilmark{6} }


\altaffiltext{1}{Academia Sinica Institute of Astronomy and Astrophysics,
P.O. Box 23-141, Taipei 106, Taiwan; cflee@asiaa.sinica.edu.tw}
\altaffiltext{2}{Graduate Institute of Astronomy and Astrophysics, National Taiwan
   University, No.  1, Sec.  4, Roosevelt Road, Taipei 10617, Taiwan}
\altaffiltext{3}{Astronomy Department, University of Virginia, Charlottesville, VA 22904, USA}
\altaffiltext{4}{INAF, Osservatorio Astrofisico di Arcetri, Largo E. Fermi 5,
50125 Firenze, Italy}
\altaffiltext{5}{Jet Propulsion Laboratory, California Institute of
Technology, Pasadena, CA 91109, USA}

\altaffiltext{6}{Harvard-Smithsonian Center for Astrophysics, 60 Garden
Street, Cambridge, MA 02138}




\begin{abstract}

HH 212 is a Class 0 protostellar system found to host a ``hamburger"-shaped
dusty disk with a rotating disk atmosphere and a collimated SiO jet at a
distance of $\sim$ 400 pc.  Recently, a compact rotating outflow has been
detected in SO and \SOO{} toward the center along the jet axis at $\sim$ 52
au (\arcsa{0}{13}) resolution.  Here we resolve the compact outflow into a
small-scale wide-opening rotating outflow shell and a collimated jet, with
the observations in the same S-bearing molecules at $\sim$ 16 au
(\arcsa{0}{04}) resolution.  The collimated jet is aligned with the SiO jet,
tracing the shock interactions in the jet.  The wide-opening outflow shell
is seen extending out from the inner disk around the SiO jet and has a width
of $\sim$ 100 au.  It is not only expanding away from the center, but also
rotating around the jet axis.  The specific angular momentum of the outflow
shell is $\sim$ 40 au \vkm{}.  Simple modeling of the observed kinematics
suggests that the rotating outflow shell can trace either a disk wind or
disk material pushed away by an unseen wind from the inner disk or
protostar.  We also resolve the disk atmosphere in the same S-bearing
molecules, confirming the Keplerian rotation there.

\end{abstract}


\section{Introduction}


HH 212 protostellar system is a well-studied young star-forming system
located in Orion at a distance of $\sim$ 400 pc.  The central source is the
Class 0 protostar IRAS 05413-0104 \citep{Zinnecker1992}.  Previous Atacama
Large Millimeter/submillimeter Array (ALMA) observations at high angular
resolution at the wavelength of 850 \micron{} have detected a vertically
resolved dusty disk feeding the central protostar \citep{Lee2017disk}.  The
disk is deeply embedded in an infalling-rotating envelope
\citep{Lee2017com}.  It has an atmosphere that shows a rotation motion
around the central star.  Detailed modeling to the kinematics of both the
envelope and disk atmosphere indicates that the disk is rotating with a
Keplerian rotation within $\sim$ 40 au of the central star.  A spinning jet
is also detected, with an inferred launching radius on the disk of $\sim$
0.05 au from the central protostar \citep{Lee2017jet}.  It can carry away
angular momentum from the innermost disk, allowing the material there to
fall onto the central protostar.

Inside the disk itself, material is expected to be transported from the
outer part to the inner part.  Thus there must be a mechanism to remove the
angular momentum from the material in the disk region outside the
jet-launching region as well.  This may be achieved with, e.g.,
magneto-rotational instability \citep{Balbus2006} and low-velocity extended
tenuous disk wind \citep{Konigl2000}.  Previous CH$_3$OH observations at
$\sim$ 240 au resolution suggested the presence of a disk wind component in
HH 212 ejected from the disk at a radius of $\sim$ 1 au \citep{Leurini2016},
surrounding the SiO jet.  However, our new observations at $\sim$ 16 au
resolution show that CH$_3$OH actually traces the disk atmosphere within a
radius of $\sim$ 40 au of the center \citep{Lee2017com}.  Nonetheless, disk
wind has also been suggested in other objects, e.g., CB 26
\citep{Launhardt2009}, DG Tau \citep{Agra-Amboage2011}, Orion BN/KL Source I
\citep{Greenhill2013,Hirota2017}, and TMC1A \citep{Bjerkeli2016}.  All these
suggest the presence of a disk wind component extracting the angular
momentum from the outer disk.



In this paper, we aim to determine whether a disk wind exists in HH 212 as
well, using our new high-resolution molecular line data.  In previous ALMA
observations at $\sim$ 200 au (\arcsa{0}{5}) resolution, \citet{Podio2015}
has detected a collimated outflow in SO and \SOO{} extending out from the
disk, inside the large-scale molecular outflow detected in C$^{34}$S
\citep{Codella2014}.  Recent follow-up observations at $\sim$ 52 au
(\arcsa{0}{13}) resolution toward the base of the collimated outflow have
detected a compact rotating outflow that may trace a disk wind
\citep{Tabone2017}.  Here, we zoom into the rotating outflow in the same
S-bearing molecules with ALMA at $\sim$ 16 au (\arcsa{0}{04}) resolution,
which is $\sim$ 3 times higher than the previous observations.  At this high
resolution, we resolve the outflow into a small-scale wide-opening rotating
outflow shell and a collimated jet aligned with the SiO jet.  We also
resolve the outflow kinematics and discuss whether the rotating outflow
shell traces a disk wind or disk material pushed away by an unseen inner
wind originated closer to the protostar.  In addition, we also resolve the
disk atmosphere in the same S-bearing molecules, confirming the Keplerian
rotation there.  Notice that some previous searches of disk wind were
done in CO \cite[see, e.g.,][]{Launhardt2009,Bjerkeli2016}.  However, here
in HH 212, CO emission is complicated, arising from both the collimated
outflow and the large-scale molecular outflow, and suffering from missing
flux around the systemic velocity.  Therefore, a detailed analysis is needed
to search for a disk wind in CO.  As a result, we defer our report of CO
observations to a future publication.


\subsection{SO and SO$_2$ around Sun-like protostars}

Sulfur is severely depleted in cold molecular clouds, with an abundance
being three orders of magnitude less than the cosmic abundance
\cite[e.g.,][]{Tieftrunk1994}.  On the other hand, S-bearing molecules are
very abundant in almost all the components associated with low-mass star
forming process, from dark clouds to protostellar envelopes and slow
outflows, as well as hot corinos \citep{vanDishoeck1998}.  This is due to
the release of sulfur from the dust mantles that can be caused by either
evaporation when the dust is heated to temperatures higher than $\sim$ 100 K
(e.g., in the hot corinos or at the disk surface), or their erosion in
shocks due to gas-grain (sputtering) or grain-grain collisions (shattering),
see e.g., \citet{Pineau1993} and \citet[and references
therein]{Guillet2011}.  In this context, S-bearing species, and in
particular, those more abundant such as SO and SO$_2$ can be considered
among the best tracers to image the high-density gas components involved in
the process leading to the formation of a Sun-like star, namely: the
high-velocity collimated jet \citep{Lee2010,Podio2015}, the accretion shock
occurring at the envelope-disk interface
\citep{Sakai2014a,Sakai2014b,Lee2016,Sakai2017}, the outflow cavity walls,
and possibly the disk surface directly illuminated and heated by the central
protostar (this paper).  The relative abundances of S-bearing molecules
depend on the main S-compounds released from mantles, i.e., if S is released
in atomic and/or molecular form such as H$_2$S or OCS, a still hotly debated
open question \citep{Wakelam2004,Codella2005,Podio2014}.  In any case,
SO$_2$ is expected to mainly form due to the reaction between SO and OH
\cite[e.g.,][]{Pineau1993}.  Indeed, the SO$_2$/SO abundance ratio has been
proposed to be an efficient chemical clock to date the gas
\citep{Charnley1997,Hatchell1998}.  Unfortunately, \citet{Wakelam2004}
and \citet{Codella2005} have shown that this abundance ratio is not very
useful, because the SO and SO2 abundances depend on the initial S-carrier on
the grain mantles.  In addition, both species may not be tracing the same
portion of the gas.


Recent observations of SO in the inner 50 au around the L1527 protostar,
located in Taurus, confirmed high SO abundances (up to 10$^{-7}$) associated
with the dense ($\ge$ 10$^{5-6}$ cm$^{-3}$) gas revealing the accretion
shock at the envelope-disk interface
\citep{Sakai2014a,Sakai2014b,Sakai2016,Sakai2017}.  On the other hand,
before the present work, \citet{Podio2015} traced SO high-velocity emission
associated with the HH 212, with very high abundances (up to 10$^{-6}$).  In
addition, the authors observed outflow-velocity emission close to the
protostar, showing small-scale velocity gradients, indicating that it
originates partly from the rotating disk, and partly from the base of the
jet, opening the way to investigations at higher spatial scale.  Also the SO
abundances in the HH 212 jet has been estimated to be very high, up to
10$^{-6}$, confirming what suggested by \citet{Lee2010} for the HH 211
protostellar jet.  Also the SO$_2$ abundance has been measured to be $\sim$
10$^{-6}$.  On the other hand, \citet{Podio2015} found in the HH 212 disk
$X_{\rm SO}$ = 10$^{-8}$--10$^{-7}$, i.e., values higher by 3--4 orders of
magnitude than those derived in protoplanetary disks
\citep{Fuente2010,Dutrey2011}.  These high abundances are plausibly due to
either violent shocks caused by jets traveling, or slow shocks occurring
close to the protostar, such as the accretion shocks already observed in HH
212 by \citet{Lee2017jet}.

\section{Observations}\label{sec:obs}


Observations of the HH 212 protostellar system were carried out with ALMA in
Band 7 at $\sim$ 350 GHz in Cycle 3, with 32-45 antennas (see Table
\ref{tab:obs}).  Two executions were carried out in 2015, one on November 5
and the other on December 3.  The projected baselines are 17-16196 m.  The
maximum recoverable size (MRS) scale is $\sim$ \arcsa{0}{4}, enough to cover
the compact outflow seen in SO and \SOO{} near the central source
\citep{Tabone2017}.  One pointing was used to map the center of the system. 
The correlator was set up to have 6 spectral windows, with one for SO \SOt{}
at 346.528481 GHz, one for CO $J=3-2$ at 345.795991 GHz, one for
H$^{13}$CO$^+$ $J=4-3$ at 346.998338 GHz, one for SiO $J=8-7$ at 347.330631
GHz, one for HCO$^+$ $J=4-3$ at 356.734288 GHz, and one for the continuum at
358 GHz with many weak \SOO{} lines (see Table \ref{tab:corr3}).  The total
time on the HH 212 system is $\sim$ 88 minutes.



The data were calibrated with the CASA package  (version 4.5) for the
passband, flux, and gain (see Table \ref{tab:calib}).  In this paper, we
only present the observational results in SO and \SOO{} in order to search
for a disk wind.  The velocity resolution is $\sim$ 0.212 \vkm{} per channel
for the lines in the spectral line windows and $\sim$ 0.848 \vkm{} per
channel for the lines in the continuum window.  In order to improve the
signal to noise ratio in the spectral line windows, we binned 4 channels to
make a bigger channel with a width of $\sim$ 0.848 \vkm{}.  We used a
Briggs weighting with a robust factor of 0.5 for the visibility weighting
to generate the SO and \SOO{} channel maps at an angular resolution of
$\sim$ \arcsa{0}{04} and a velocity resolution of $\Delta v_c = 0.848$
\vkm{} per channel.  In generating the SO and \SOO{} maps, the data with the
$uv$-distance greater than 12000 m and 8000 m (with a corresponding angular
scale of $\sim$ \arcsa{0}{015} and \arcsa{0}{022}, respectively) are
excluded respectively because no emission is detected there.  Also an
outer taper of \arcsa{0}{03} was used to obtain a better signal-to-noise
ratio. The noise level can be measured from line-free channels.  It is
$\sim$ 1.85 \mJyb{} (or $\sim$ 12.4 K) for the SO channel maps.  Twelve
\SOO{} lines are detected (see Table \ref{tab:lines}) and stacked, and the
noise level is $\sim$ 0.48 \mJyb{} (or $\sim$ 3.0 K) for the resulting
\SOO{} channel maps.  The velocities in the channel maps are LSR velocities. 
Various integrated maps are produced with various velocity ranges in
order to show the low-velocity, high-velocity, and total integrated maps. 
The rms levels in those maps are estimated from the emission-free region in
the same map or the emission-free integrated map with the same velocity
width.






\section{Results}





\begin{figure} [!hbp]
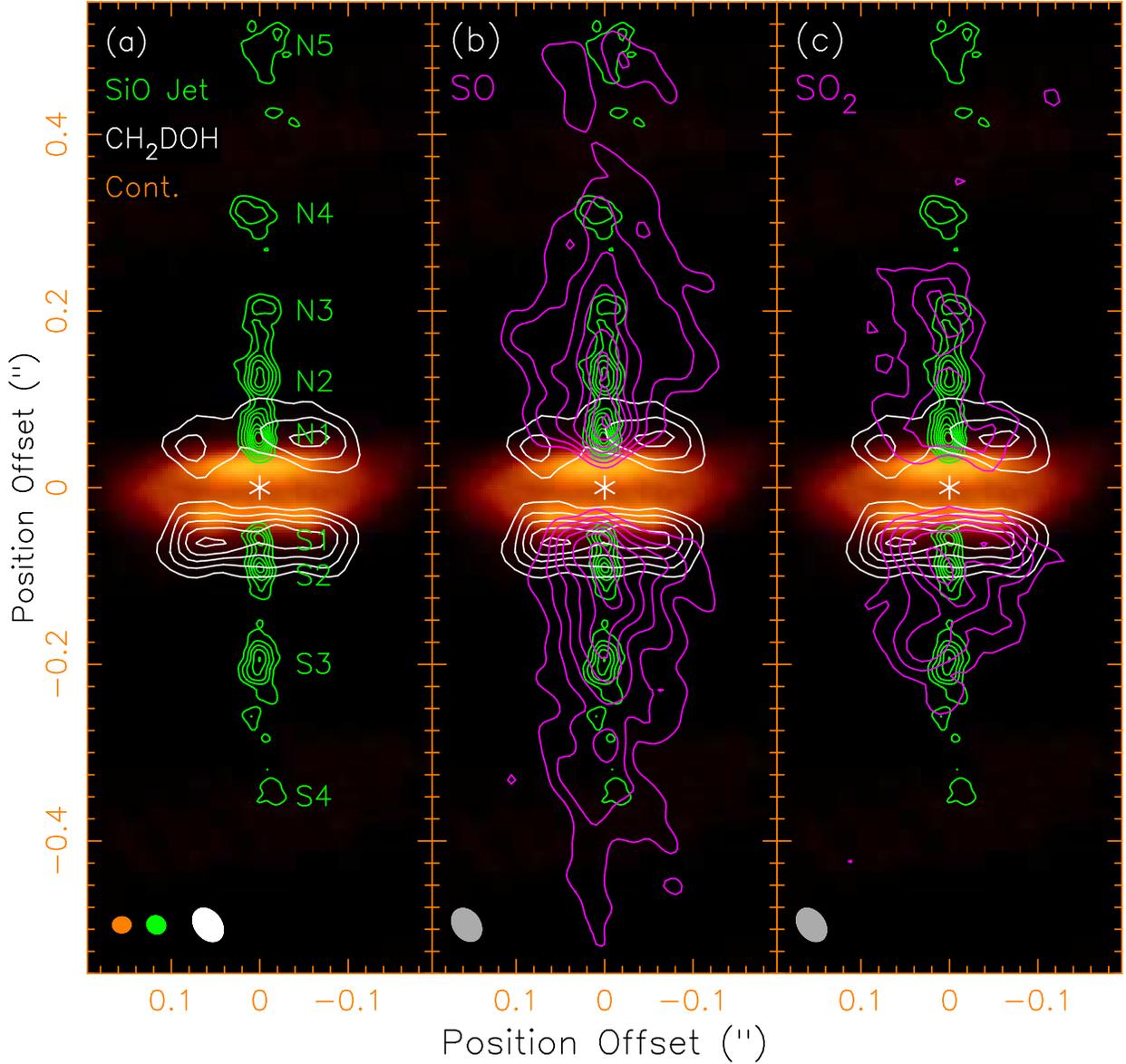
 \centering \putfig{0.9}{270}{f1.eps}
\figcaption[] {ALMA SO and \SOO{} maps toward the center
of the HH 212 system, on top of the dusty disk \cite[orange image,
from][]{Lee2017disk}, disk atmosphere \cite[white contours,
from][]{Lee2017com}, and the SiO jet \cite[green contours,
from][]{Lee2017jet}.  The maps are all rotated by 22.5\degree{} clockwise to
align the jet axis in the north-south direction.  The star marks the
position of the central protostar.  The magenta contours in (b) and (c) show
the total integrated maps of SO ($\Voff \sim$ $-$10.0 to 10.0 \vkm) and
\SOO{} ($\Voff \sim$ $-$7.9 to 6.5 \vkm), respectively.  The contours start
at 3 $\sigma$ with a step of 2 $\sigma$, where $\sigma$= 0.009 \Jybk{} and
0.0022\Jybk{} for SO and \SOO{}, respectively.  \label{fig:SO_SO2}}
\end{figure}

\begin{figure} [!hbp]
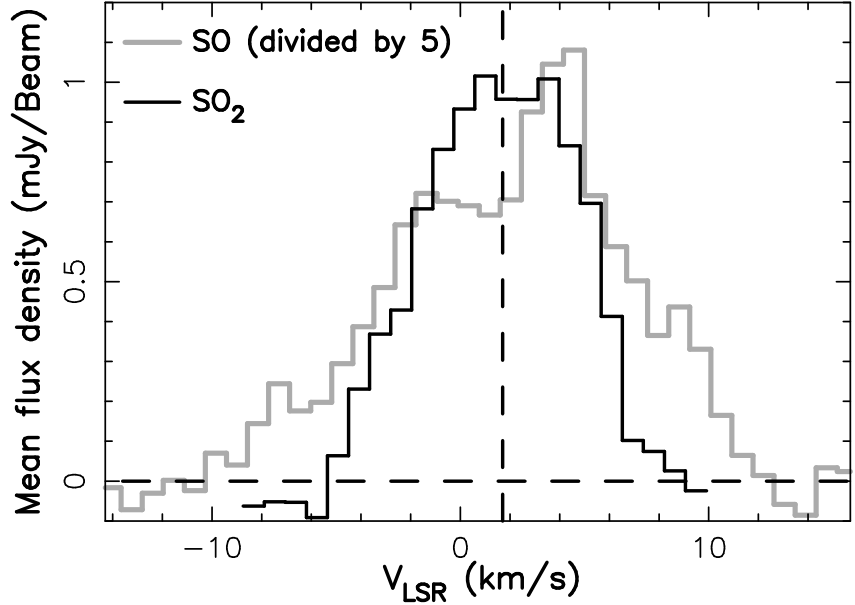

\centering
\putfig{0.9}{270}{f2.eps} %
\figcaption[]
{SO (gray curve) and \SOO{} (black curve) spectra averaged over a rectangular region of
\arcsa{0}{4}$\times$\arcsa{0}{2} centered at the central source aligned with the jet axis.
The SO flux density has been divided by 5 in order to be plotted with SO$_2$.
\label{fig:spec}}
\end{figure}

For comparison, Figure \ref{fig:SO_SO2}a shows the previous maps of dusty
accretion disk in 850 \micron{} continuum \citep{Lee2017disk}, disk
atmosphere in \CHtDOH{} \citep{Lee2017com}, and collimated jet in SiO
\citep{Lee2017jet}.  These maps have been rotated by 22.5\degree{} clockwise
to align the jet axis in the north-south direction in order to facilitate
our presentations.  The jet has a mean inclination angle of $\sim$
4\degree{}$\pm$2\degree{} to the plane of the sky, with the northern
component tilted toward us \citep{Claussen1998,Lee2007}.  The disk is
perpendicular to the jet axis and nearly edge-on with the nearside titled
slightly by $\sim$ 4\degree{}$\pm$2\degree{} to the south
\citep{Lee2017disk}.  The atmosphere is better detected in the south than in
the north.  In the following, our maps of SO and \SOO{} are presented
together with the jet, disk, and disk atmosphere in order to show the
connection to them.  In this system, the systemic velocity is assumed to be
$\Vsys= 1.7\pm0.1$ \vkm{} LSR \citep{Lee2007}, which is a mean value
estimated before in CO J=1-0 \citep{Lee2000} and NH3 (1,1)
\citep{Wiseman2001}.  Figure \ref{fig:spec} shows our SO and \SOO{} spectra
extracted from a rectangular region (\arcsa{0}{4}$\times$\arcsa{0}{2} with
the long side aligned with the jet axis) covering the emission from the
central region.  As can be seen, the systemic velocity lies roughly at the
center of the spectra.  Throughout this paper, in order to facilitate our
presentations, we define an offset velocity $\Voff = \VLSR - \Vsys$.  Also,
velocities with $|\Voff| \lesssim$ 3 \vkm{} are referred to as low and those
outside the range as high.  The low-velocity range is selected to show the
disk atmosphere and a small-scale wide-opening outflow extending out from
the inner disk, as discussed below.


\subsection{SO Emission Morphology} \label{sec:SO}



A compact outflow has been detected recently in SO at $\sim$ \arcsa{0}{13}
resolution, extending out from the central source along the jet axis
\citep{Tabone2017}.  Now at $\sim$ \arcsa{0}{04} resolution, we can resolve
the outflow and check for a disk wind component.  As shown in the total
integrated emission map in Figure \ref{fig:SO_SO2}b, the SO emission is
detected here within $\sim$ $\pm$\arcsa{0}{5} of the central source around
the SiO jet, extending out from the inner disk.  The emission has a velocity
with $|\Voff| \lesssim$ 10 \vkm{}.  SO emission is also detected in the disk
atmosphere, overlapping with the \CHtDOH{} emission.  No SO emission is
detected in the dusty disk, likely because the continuum emission is bright
and optically thick there \citep{Lee2017com}; thus, the emission behind
it is blocked and the emission in front of it appears absorbed against the
bright background.





\begin{figure} [!hbp]
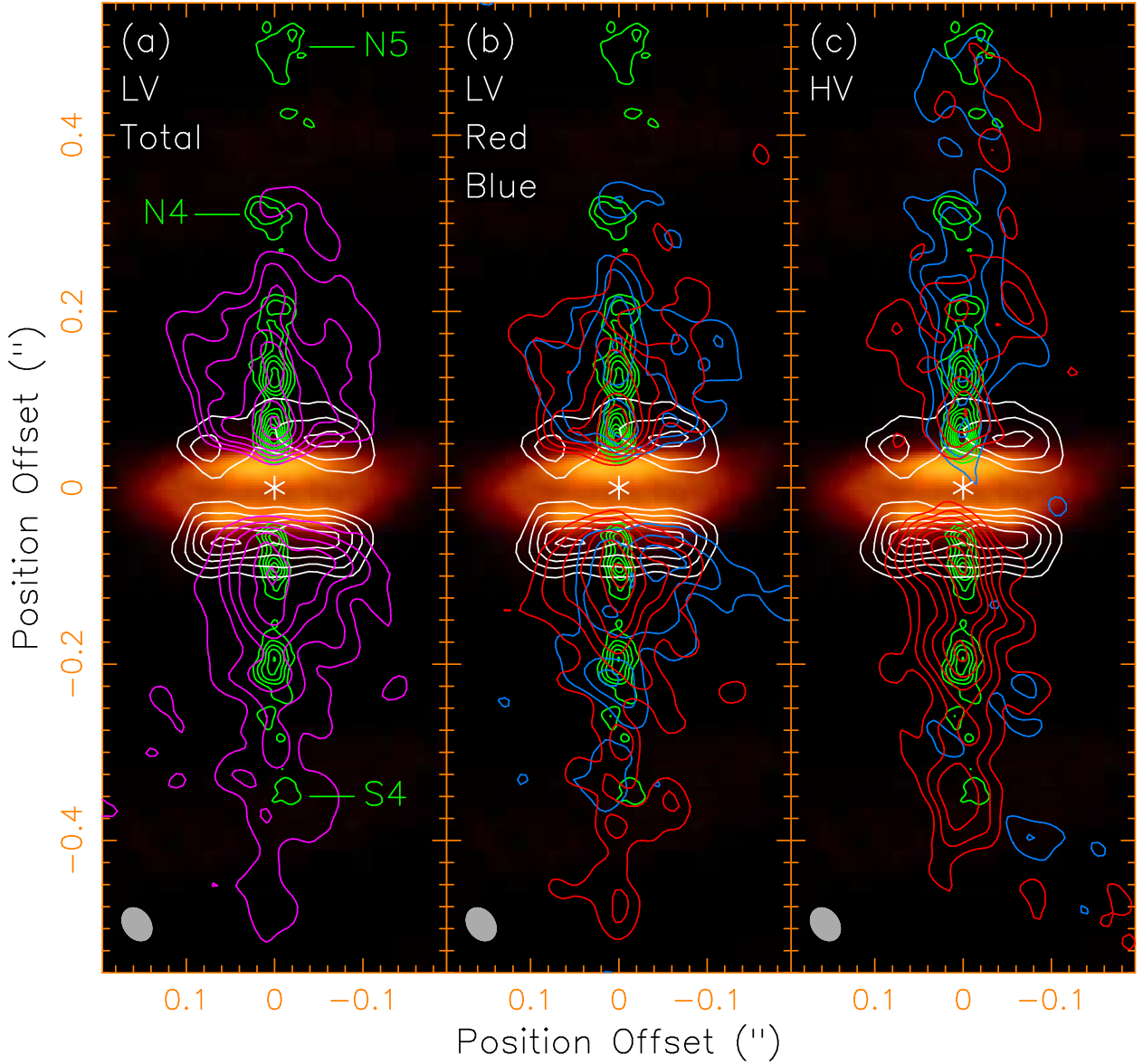

\centering
\putfig{0.9}{270}{f3.eps} 
\figcaption[]
{Low-velocity and high-velocity SO maps. 
The orange image, white contours, and green contours have the same meaning
as in Figure \ref{fig:SO_SO2}.
(a) shows the total low-velocity map
with $\Voff\sim$-3 to 3 \vkm{}.
(b) shows the blueshifted ($\Voff\sim$-3 to 0 \vkm{})
and redshifted ($\Voff\sim$0 to 3 \vkm{}) low-velocity maps separately,
(c) shows the blueshifted ($\Voff\sim$-10 to -3 \vkm{}) and redshifted 
($\Voff\sim$ 3 to 10 \vkm{})
high-velocity maps separately.
The star marks the protostar position. The white curves in (a) guide
the readers for the shell structure.
The contour levels start at 3 $\sigma$ with a step of 2$\sigma$, where
$\sigma=$ 4.5 \mJybk{} in (a) and (c), and 3 \mJybk{} in (b).
\label{fig:so}}
\end{figure}

Figure \ref{fig:so} shows the SO maps at low and high velocities.  At low
velocity, the outflow can be decomposed into two components, a wide-opening
outflow extending out to $\sim$ \arcsa{0}{2} to the north and south from the
inner disk, and a collimated jet along the jet axis (Figure \ref{fig:so}a). 
The wide-opening outflow is reasonably resolved in the north, appearing as a
shell structure opening out from the central source.  The shell structure of
the outflow can be better seen in the position-velocity (PV) diagrams as
discussed below in Section \ref{sec:kin}.  In the shell outside the
atmosphere, the blueshifted emission overlaps with the redshifted emission
(Figure \ref{fig:so}b), indicating that the shell there is mainly radially
expanding.  At high velocity,  the SO emission is seen around and along
the jet axis (Figure \ref{fig:so}c).  As discussed later, since the
projected velocity of the expanding shell can reach up to $\Voff$ of $\pm$5
\vkm{}, part of the emission traces the expanding shell projected on the jet
axis.  The emission peaks that spatially coincide with knots N3, N1, S1/S2,
S3 and S4 should mainly trace the jet itself. Like the SiO jet, the
blueshifted emission is seen more to the north and the redshifted
emission is seen more to the south.


\subsection{\SOO{} Emission Morphology} \label{sec:SOt}

A compact outflow has also been detected recently in \SOO{} at $\sim$
\arcsa{0}{13} resolution \citep{Tabone2017}.  Thus, we can also check for a
disk wind component in \SOO{}.  Here twelve \SOO{} lines are detected (see
Table \ref{tab:lines}) with $|\Voff| \lesssim$ 8 \vkm{}.  They are weak and
thus stacked to produce a mean \SOO{} line with a better signal to noise
ratio.  As shown in the total integrated emission map in Figure
\ref{fig:SO_SO2}c, the \SOO{} emission is detected within $\sim$
$\pm$\arcsa{0}{3} of the central source, not only tracing the outflow but
also tracing the inner disk atmosphere.  No \SOO{} emission is detected
toward the dusty disk because of the same reason as discussed above.



\begin{figure} [!hbp]
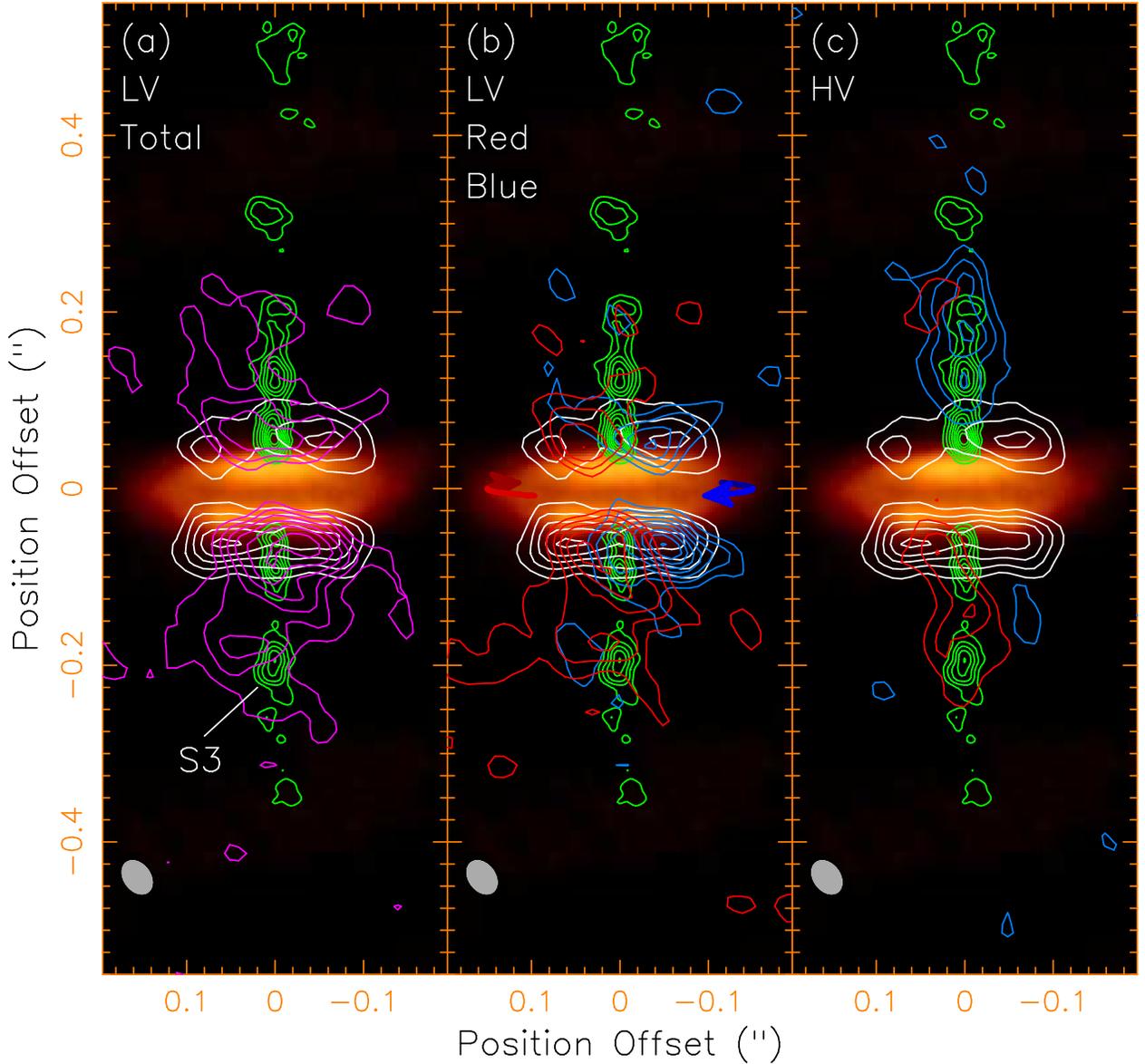

\centering
\putfig{0.9}{270}{f4.eps} 
\figcaption[]
{Low-velocity and high-velocity \SOO{} maps. 
The orange image, white contours, and green contours have the same meaning
as in Figure \ref{fig:SO_SO2}.
(a) shows the total low-velocity map,
with $\Voff \sim$ -3 to 3 \vkm{}.
(b) shows the blueshifted ($\Voff \sim$ -3 to 0.0 \vkm{})
and redshifted ($\Voff \sim$ 0 to 3 \vkm{})
low-velocity maps separately.
The red and blue arrows show
the rotation of the disk.
(c) shows the blueshifted ($\Voff \sim$ -7.9 to -3 \vkm{}) and 
redshifted ($\Voff \sim$ 3 to 6.5 \vkm{}) high-velocity maps separately.
The star marks the protostar position. The white curves in (a) guide
the readers for the shell structure.
The contour levels start at 3 $\sigma$ with a step of 2$\sigma$, where
$\sigma=$ 1.5 \mJybk{} in (a), and 1.1 \mJybk{} in (b) and (c).
\label{fig:so2}}
\end{figure}

Figure \ref{fig:so2} shows the \SOO{} maps at low and high velocities.  At
low velocity, unlike the SO emission, the \SOO{} emission traces mainly the
inner disk atmosphere, as seen in Figures \ref{fig:so2}a and \ref{fig:so2}b. 
The disk atmosphere is rotating with the redshifted emission in the east and
blueshifted emission in the west, the same rotation sense as that found
before in \CHtDOH{}, as indicated by the blue and red arrows in Figure
\ref{fig:so2}b.  Like that seen in SO, the outflow in the north can also be
decomposed into a wide-opening outflow and a collimated jet.  In the
north, the shell structure of the wide-opening outflow can be seen in the
east.  In the south, the shell structure can only be barely seen in the
east, because it is contaminated by the emission around knot S3.  At high
velocity, like the SO emission, the \SOO{} emission is mainly seen around
the jet axis, with the blueshifted emission to the north and redshifted
emission to the south (Figure \ref{fig:so2}c).

\subsection{Kinematics}\label{sec:kin}

\begin{figure} [!hbp]
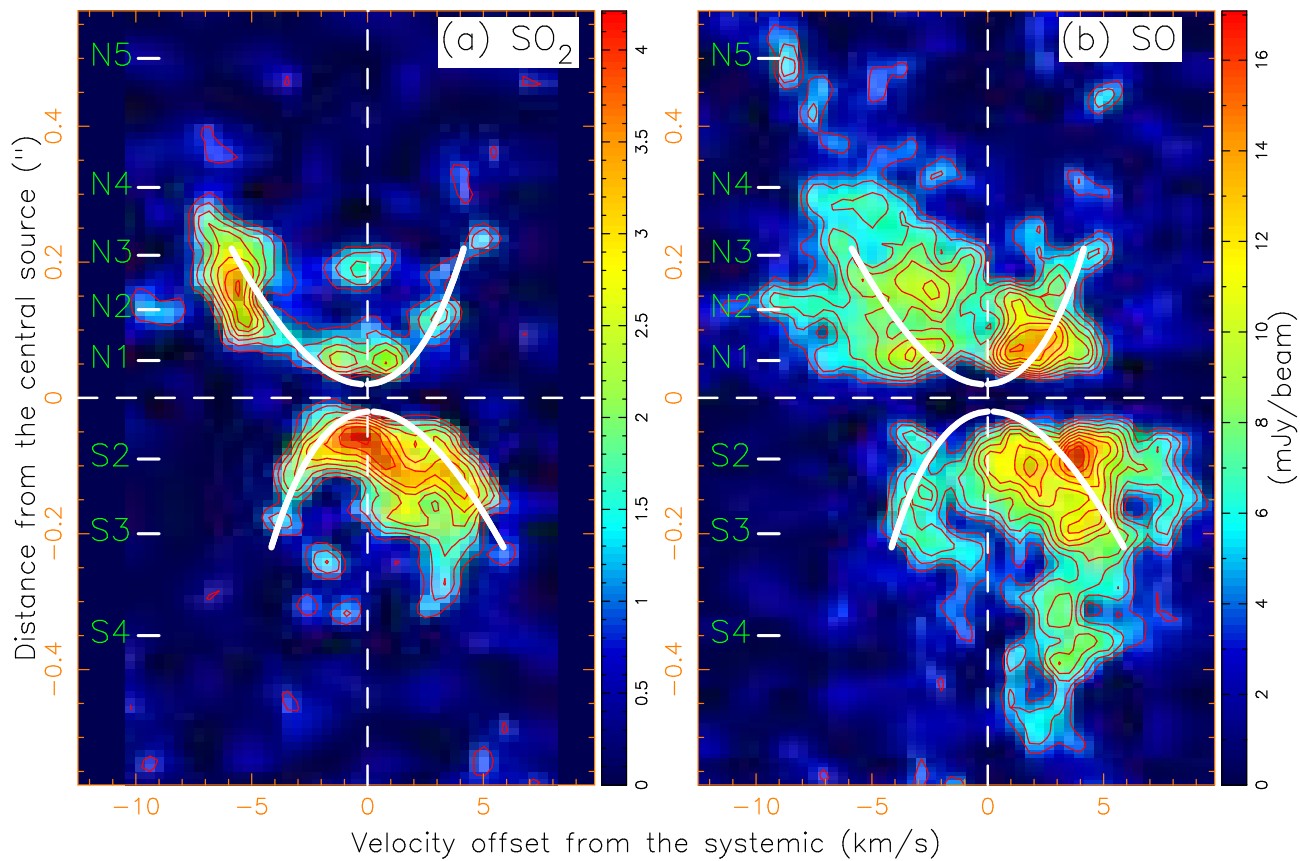

\centering
\putfig{0.65}{270}{f5.eps} 
\figcaption[]
{Position-Velocity diagrams of the SO (contours) and \SOO{} (color image)
emissions cut along the jet axis. The contours start
at 3 $\sigma$ with a step of 1 $\sigma$, where
$\sigma=$ 1.4 \mJyb{}. The white curves guide the readers for the parabolic PV structure
of the outflow shell.
\label{fig:pvjet}}
\end{figure}

Figure \ref{fig:pvjet} shows the position-velocity (PV) diagrams along the
jet axis in SO and \SOO{} in order to study the expansion velocity of the
wide-opening outflow.  The PV diagram of the \SOO{} emission shows roughly a
parabolic structure opening up from the source position to $\sim$
\arcsa{0}{2} to the north and south to $\Voff \sim \pm5$ \vkm{}, as guided
by the white curves.  The PV diagram of the SO emission is complicated. 
Part of the SO emission follows the parabolic structure seen in \SOO{} and
part of the SO emission is associated with the knots in the jet.  The
parabolic PV structure is associated with the wide-opening outflow,
supporting that the wide-opening outflow is a shell and that the expansion
velocity of the shell increases with the distance.

\begin{figure} [!hbp]
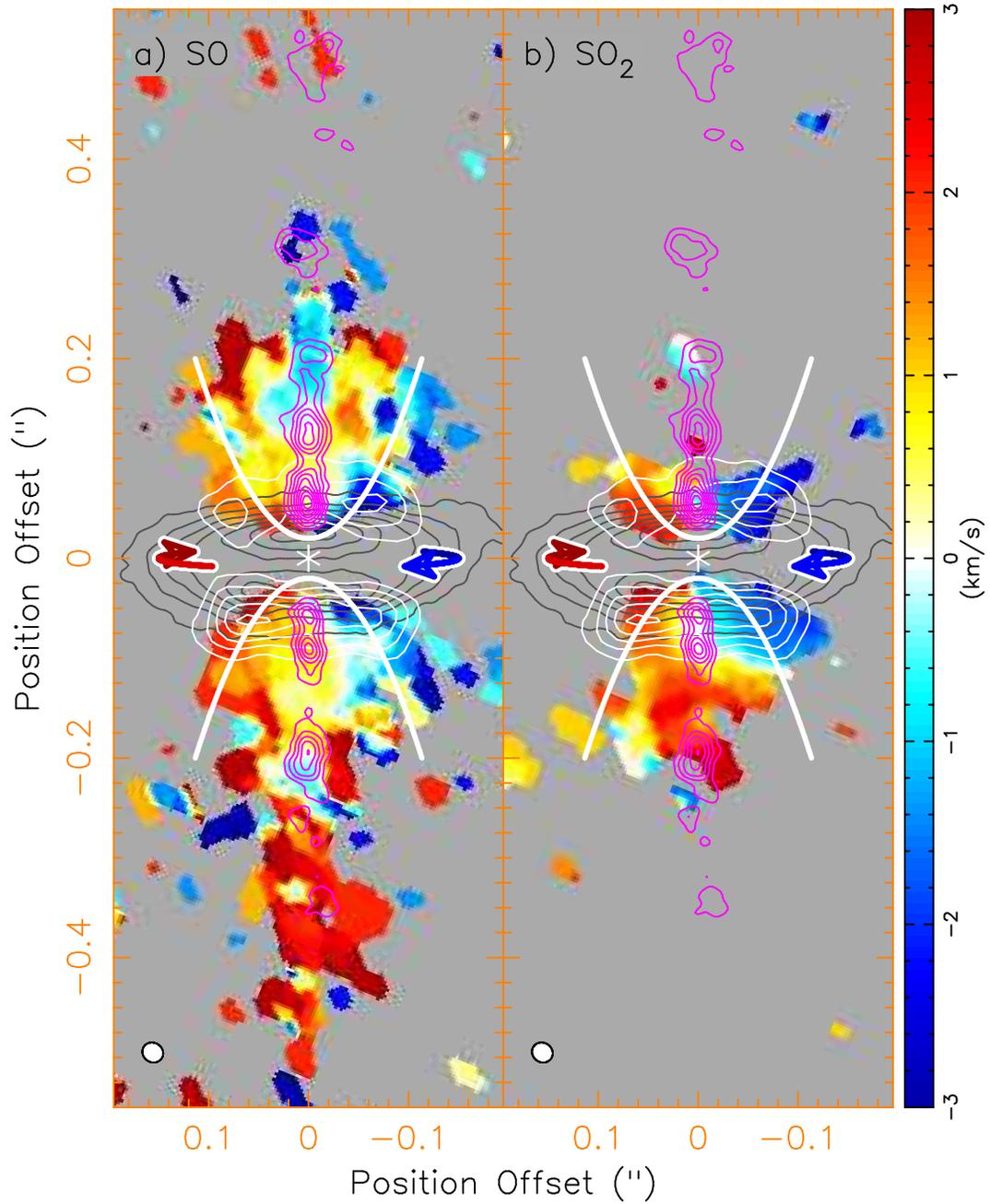

\centering
\putfig{1}{270}{f6.eps} 
\figcaption[]
{The intensity weighted velocity (first moment)
maps of the SO and \SOO{} emission at low velocities, plotted together
with the maps of the dusty disk (gray contours), disk atmosphere (white contours),
and SiO jet (magenta contours). The red and blue arrows show
the rotation of the disk.
\label{fig:mom1}}
\end{figure}

Figure \ref{fig:mom1} shows the intensity weighted velocity (first moment)
maps of the SO and \SOO{} emission at low velocities in order to search for
the rotation in the shell.  In SO, the eastern and western outer edges
(as roughly delineated by the white curves) of the shell mostly have
opposite velocities (Figure \ref{fig:mom1}a), indicating that the shell is
also rotating around the jet axis, in addition to the expansion mentioned
above.  The rotation of the shell has the same sense as that of the disk, as
indicated by the red and blue arrows.  On the other hand, as discussed
earlier, the \SOO{} emission mainly arises from the rotating disk atmosphere
(Figure \ref{fig:mom1}b).  The outflow edges are too faint in \SOO{} to
appear in this figure.






\subsubsection{Rotation and Expansion in the Shell}

\begin{figure} [!hbp]
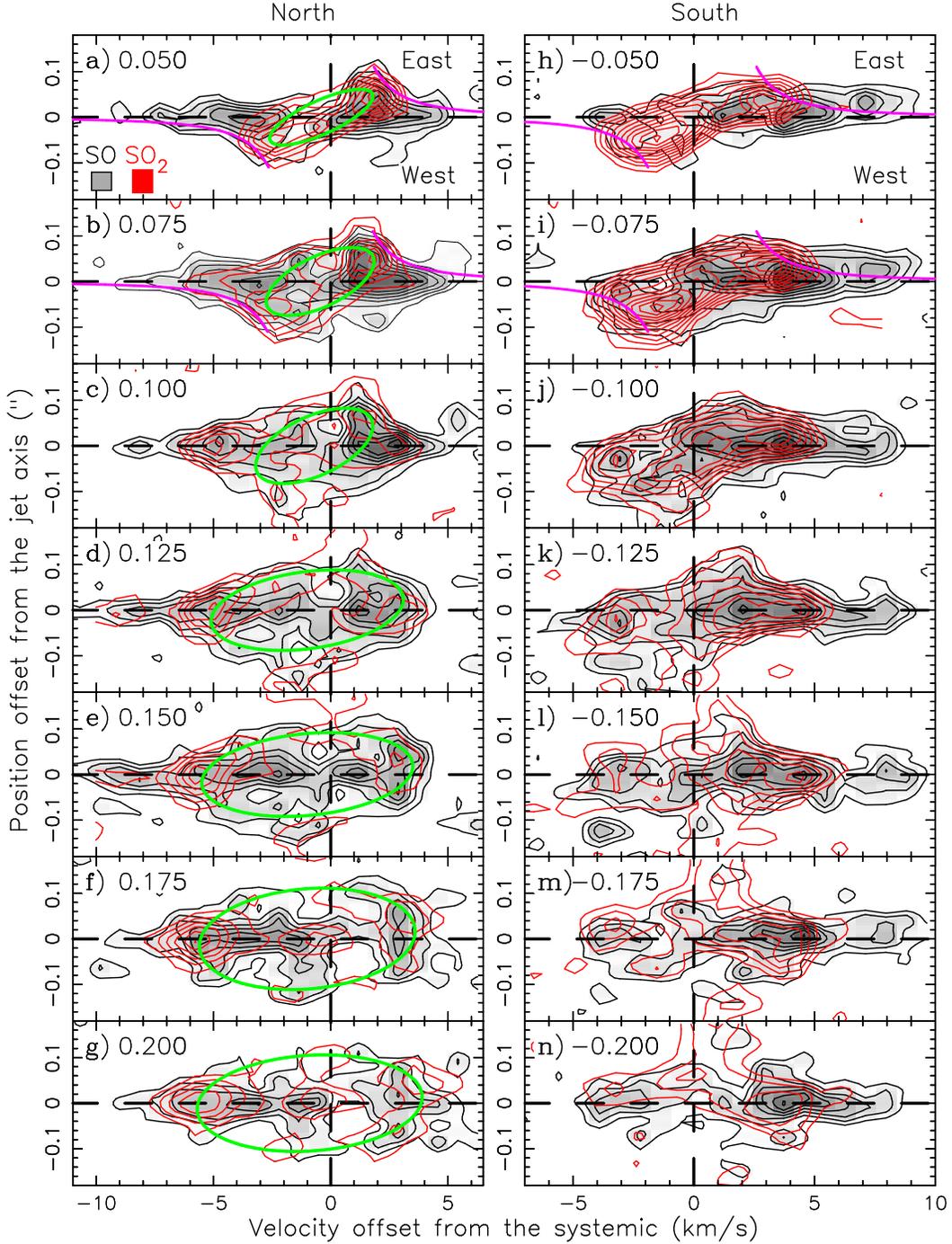
 \centering \putfig{0.75}{0}{f7.eps} 
\figcaption[] {PV diagrams of the SO and \SOO{} emissions across the jet
axis at increasing distance from the central source.  Left column for the
north and right for the south.  The number in the upper left corner
indicates the distance from the central source along the jet axis.  Gray
image with black contours is for SO emission.  The contours start at
3$\sigma$ with a step of 2 $\sigma$, where $\sigma=1.1\times10^{-3}$ \Jyb{}. 
Red contours are for \SOO{} emission.  The contours start at 3$\sigma$ with
a step of 1.5 $\sigma$, where $\sigma=3.4\times10^{-4}$ \Jyb{}.  The green
ellipses guide the readers for the elliptical PV structures of the shell. 
The magenta curves show the Keplerian rotation velocity profile due to the
central protostar \citep{Lee2017com}.  In the upper left panel, the
rectangles in the lower left corner show the resolutions of the cuts in SO
and \SOO{}.  \label{fig:pvobs}} \end{figure}

Figure \ref{fig:pvobs} shows the PV diagrams cut across the jet axis at
increasing distance $z$ from the disk midplane with a step of \arcsa{0}{025}
(or 10 au) to the north and south, in order to study the expansion and
rotation in the shell in detail.  Since the kinematics in the northern part
is better resolved than that in the southern part, our kinematic study here
is based mainly on the northern part.

Near the bottom of the northern atmosphere at $z=$ \arcsa{0}{05}, the PV
diagram shows the redshifted emission in the east and blueshifted emission
in the west (Figure \ref{fig:pvobs}a), as expected for a rotating disk
atmosphere.  In addition, the SO emission shows a velocity increasing toward
the center, with the outer boundary of its PV structure well described by
the Keplerian rotation velocity profile (magenta curves) due to the central
protostar \citep{Lee2017com}. This confirms the presence of a rotationally
supported disk orbiting the protostar. The mass of the protostar
is $\sim$ 0.25 \solarmass{} from the kinematics of the disk and envelope
in molecular lines \citep{Lee2017com}. At low velocities where the emission
traces the outer part of the atmosphere, the PV structure can be roughly
described by a tilted elliptical structure as guided by the green ellipse.
This outer part is also seen in \SOO{} emission.  This PV structure confirms
that the outer part of the atmosphere is not only rotating but also
expanding \citep{Hirota2017}, indicating that the atmosphere there has
become a part of the outflow shell.  The upper part of the ellipse is much
fainter than the lower part, probably because the emission in the nearside
is self-absorbed.  As we go higher up in the atmosphere to $z=$
\arcsa{0}{075}, the elliptical PV structure becomes broader in the minor
axis (Figure \ref{fig:pvobs}b), indicating that the expansion motion becomes
faster higher up, consistent with the discussion above.

In the shell slightly above the disk atmosphere at $z=$ \arcsa{0}{1}, the PV
diagrams also show a tilted elliptical PV structure (green ellipse, Figure
\ref{fig:pvobs}c), indicating that the shell there is also rotating and
expanding.  As compared to that seen in the atmosphere, it is broader in the
minor axis, indicating that the expansion motion is faster.  As we go
further to the north to $z\geq$ \arcsa{0}{125}, the PV structure can be
decomposed into two components, one with an elliptical structure for the
shell, and the other along the jet axis tracing the emission in the jet
(Figures \ref{fig:pvobs}d-\ref{fig:pvobs}g).  The major axis of the ellipse
is almost aligned with the x-axis, indicating that the gas motion in the
shell is dominated by the expansion.  In addition, the \SOO{} emission of
the shell becomes very faint at low velocities, thus the limb-brightened
shell (outflow edge) becomes faint and not detected, as seen in Figure
\ref{fig:so2}a.  As a result, the emission of the shell is mainly detected
at high velocities, appearing mainly around the jet axis, as seen in Figure
\ref{fig:so2}c.

In the southern part, similar elliptical PV structures can also be roughly
seen for the shell in the atmosphere (Figures \ref{fig:pvobs}h and
\ref{fig:pvobs}i).  However, no clear elliptical PV structures can be
identified easily (Figures \ref{fig:pvobs}j to \ref{fig:pvobs}n) in the PV
diagrams above the disk atmosphere.  Therefore, we will not attempt to
fit the PV structures here because the results would be quite uncertain.





\begin{figure} [!hbp]
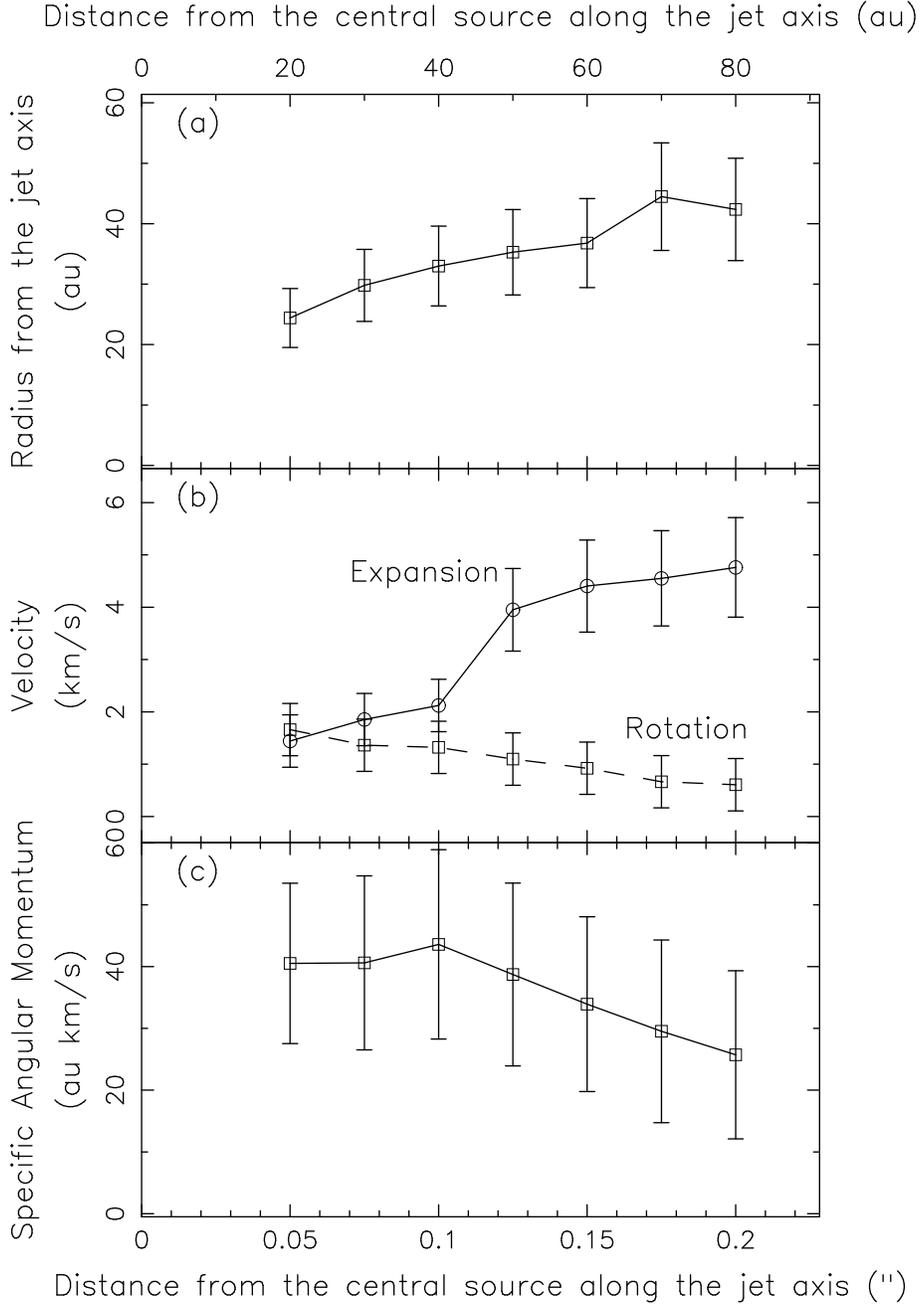

\centering
\putfig{0.7}{0}{f8.eps} 
\figcaption[]
{Radius, expansion velocity, rotation velocity, and specific angular momentum
of the shell in the north at different distance from the central source 
above the disk midplane. They are derived from the elliptical PV structures
seen in Figures \ref{fig:pvobs}a-g. 
The radius is assumed to
have an uncertainty of 20\%.  The expansion velocity is assumed to have an
uncertainty of 20\% or 0.5 \vkm{} (about a half of the channel width),
whichever is larger.  The rotation velocity is small and thus assumed to
have an uncertainty of 0.5 \vkm{} (i.e., about a half of the channel width). 
The resulting uncertainty in the specific angular momentum is about
30\%-50\%. 
\label{fig:vexprot}}
\end{figure}

The physical properties (e.g., radius, rotation velocity, expansion
velocity, and specific angular momentum) of the shell in the north, where
the shell is detected at a higher signal-to-noise ratio and better resolved,
can be estimated from the elliptical PV structures seen in the PV diagrams
cut across the jet axis in Figures \ref{fig:pvobs}a-g.  As can be seen, the
elliptical PV structures can be roughly described by the green ellipses in
the figures.  Since the elliptical PV structures are not well defined, we
only used an eye-fitting to roughly obtain the green ellipses, which are
required to pass through the emission in the shells as much as possible. 
From these green ellipses, the radius of the shell is given by the maximum
position offset from the jet axis, the rotation velocity is given by the
velocity at the maximum position offset with respect to the velocity
centroid of the ellipse, and the expansion velocity is given by the velocity
at the zero position offset with respect to the velocity centroid of the
ellipse.  The specific angular momentum is then derived by multiplying the
rotation velocity with the radius.  These physical quantities of the
shell are shown in Figure \ref{fig:vexprot}, with their uncertainties
described in the caption.
 Since the inclination angle of the outflow is small, no correction is
needed for the inclination effect.  As expected, the shell becomes wider
with the distance (Figure \ref{fig:vexprot}a).  Interestingly, the rotation
velocity is roughly the same as the expansion velocity in the shell inside
the atmosphere, but becomes much smaller than the expansion velocity in the
shell outside the atmosphere (Figure \ref{fig:vexprot}b).  The specific
angular momentum in the shell is roughly constant at $\sim$ 40 au \vkm{}
within $\sim$ 50 au of the central source, and then decreases to $\sim$ 25
au \vkm{} at $\sim$ 80 au (Figure \ref{fig:vexprot}c).  This decrease in the
specific angular momentum is uncertain because the rotation velocity can not
be estimated accurately outside the atmosphere due to the poorly identified
elliptical PV structure.


\subsection{Physical condition in the disk atmosphere and
shell}\label{sec:colden}

With twelve \SOO{} lines detected, we can construct a population diagram
\citep{Goldsmith1999} to estimate the rotational temperature and the column
density of the molecule in the disk atmosphere, the wide-angle outflow
shell, and the collimated jet.  It is a diagram that plots the column
density per statistical weight in the upper energy state in the optically
thin limit, $N_u^\textrm{\scriptsize thin}/g_u$, versus the upper energy
level $E_u$ of the lines.  Here $N_u^\textrm{\scriptsize thin}=(8\pi
k\nu^2/hc^3 A_{ul}) W$, where the integrated line intensity $W=\int T_B dv $
with $T_B$ being the mean brightness temperature.

\begin{figure} [!hbp]
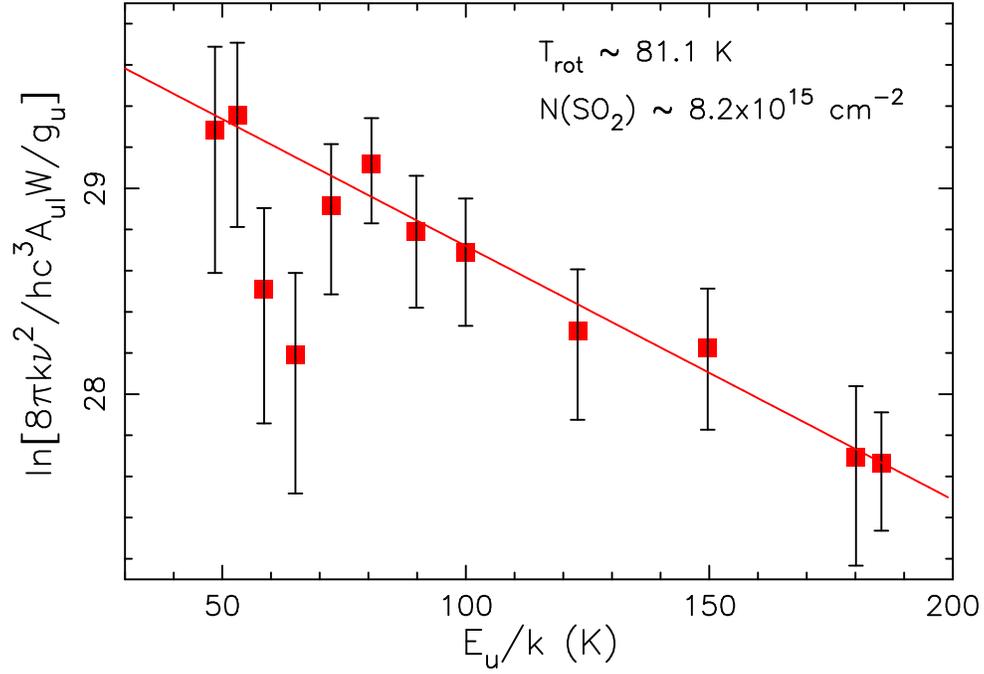

\centering
\putfig{1}{270}{f9.eps} 
\figcaption[]
{
The population diagram for \SOO{}, derived from the total intensity map
for the disk atmosphere and outflow shell in the south. 
The error bars show the uncertainties in our measurements, 
which are assumed to scale inversely with
the integrated line intensity, with 25\% of the data values for the highest integrated line intensity
and 50\% for the lowest integrated line intensity.
The solid line is a linear fit to the data with E$_u$ $>$ 70 K.
\label{fig:pop}}
\end{figure}

The integrated line intensity of each transition can be measured for the
\SOO{} emission in the south, where the emission is detected at a
higher signal-to-noise ratio.  The resulting population diagram is shown in
Figure \ref{fig:pop}.  Fitting the data points with E$_\textrm{u} > 70$ K,
we estimate a rotational temperature of 81$\pm$20 K and a column density of
8.2$\pm2.0\times10^{15}$ \cms{}.  Notice that the data points with
E$_\textrm{u} < 70$ K are not used for the fitting because their emissions
are only barely detected, with their intensity slightly over 3 $\sigma$. 
These estimated values are mainly for the disk atmosphere and the outflow
shell, because \SOO{} emission is mainly detected there.



The SO emission traces mainly the outflow shell and the  collimated
jet.  The excitation temperature is unknown.  Along the jet axis, the
emission has a peak brightness temperature of $\sim$ 150 K and a mean
integrated line intensity of 700 K \vkm{}.  Thus, we assume a temperature of
$\sim$ 200-300 K in the collimated jet, consistent with the
temperature range discussed in \citet{Podio2015}.  The column density of SO
there is then estimated to be $\sim$ 1.5-1.8$\times10^{16}$ \cms{}.  In the
shell, the mean integrated line intensity is $\sim$ 400 K \vkm{}.  The
excitation temperature can be assumed to be $\sim$ 100 K, which is an
approximation to that derived above in \SOO{} for the disk atmosphere and
the outflow shell.  Thus, the SO column density is estimated to be $\sim$
6$\times10^{15}$ \cms{} in the outflow shell.

The abundances of SO and \SOO{} have been estimated before at lower
resolution by comparing their column density to that derived from the CO
J=3-2 emission \cite[see Table 3 in][]{Podio2015} and thus can be used here
to derive the mean H$_2$ column density.  With a SO abundance of
$(1-10)\times10^{-7}$, the mean H$_2$ column density is estimated to be
$(6-60)\times10^{21}$ \cms{} in the SO shell.  With a \SOO{} abundance of
$(5-12)\times10^{-7}$, the mean H$_2$ column density is estimated to be
$(5-12)\times10^{21}$ \cms{} in the \SOO{} shell.  Given that the thickness
of the shell is $\lesssim$ 50 au, the mean H$_2$ density is $\gtrsim$
$(4-40)\times10^{6}$ \cmc{} in the SO shell, and $(3-8)\times10^{6}$ \cmc{}
in the \SOO{} shell.  As can be seen, the SO shell could have a higher mean
H$_2$ density than that in the \SOO{} shell.




\section{Discussion}

\subsection{Disk wind?}


In star formation, one important topic is to search for a low-velocity disk
wind that can carry away the disk angular momentum, allowing the disk
material to accrete.  Previously, a rotating outflow was detected in CB 26
in CO J=2-1 emission and was argued to trace a disk wind launched from a
radius of $\sim$ 30 au in the disk \citep{Launhardt2009}.  Later, a rotating
outflow was also detected in Orion BN/KL Source I in SiO maser, appearing to
be a disk wind launched from a radius of $\sim$ 20 au in the disk
\citep{Matthews2010}.  In recent follow-up observations of the same object,
a rotating outflow was also detected further out in Si$^{18}$O, and argued
to trace a disk wind launched from a radius of $>$ 10 au \citep{Hirota2017}. 
Recently, another rotating outflow is detected in TMC1A in CO J=2-1, and
also argued to trace a disk wind launched from a radius of $>$ 5 au in the
disk \citep{Bjerkeli2016}.  Another example is HH 30, where a rotating CO
(J=2-1) shell is observed to surround one side of the bipolar optical jet
(Louvet et al.  in prep).

In HH 212, we can search for a disk wind emerging from the well-resolved
disk.  As discussed later, the wind's source can be on the disk 2-3 au from
the protostar.  At this radius, the density and temperature on the disk
surface (at one scale height) are estimated to be $\sim 2-4\times10^{12}$
\cmc{} and 770-1050 K, respectively, based on the disk model in
\citet{Lee2017disk}.  Thus, our SO and \SOO{} lines, with a critical density
of $\sim 10^7$ \cmc{}, are better tracers of a disk wind than the CO J=2-1
line, which has a critical density of $\sim$ $10^4$.  Note that however,
since only charged particles are directly coupled to the magnetic field,
neither CO and SO/SO2 can be directly accelerated by a magnetic field; they
must be coupled to magnetically accelerated charged particles through
collisions.

\begin{figure} [!hbp]
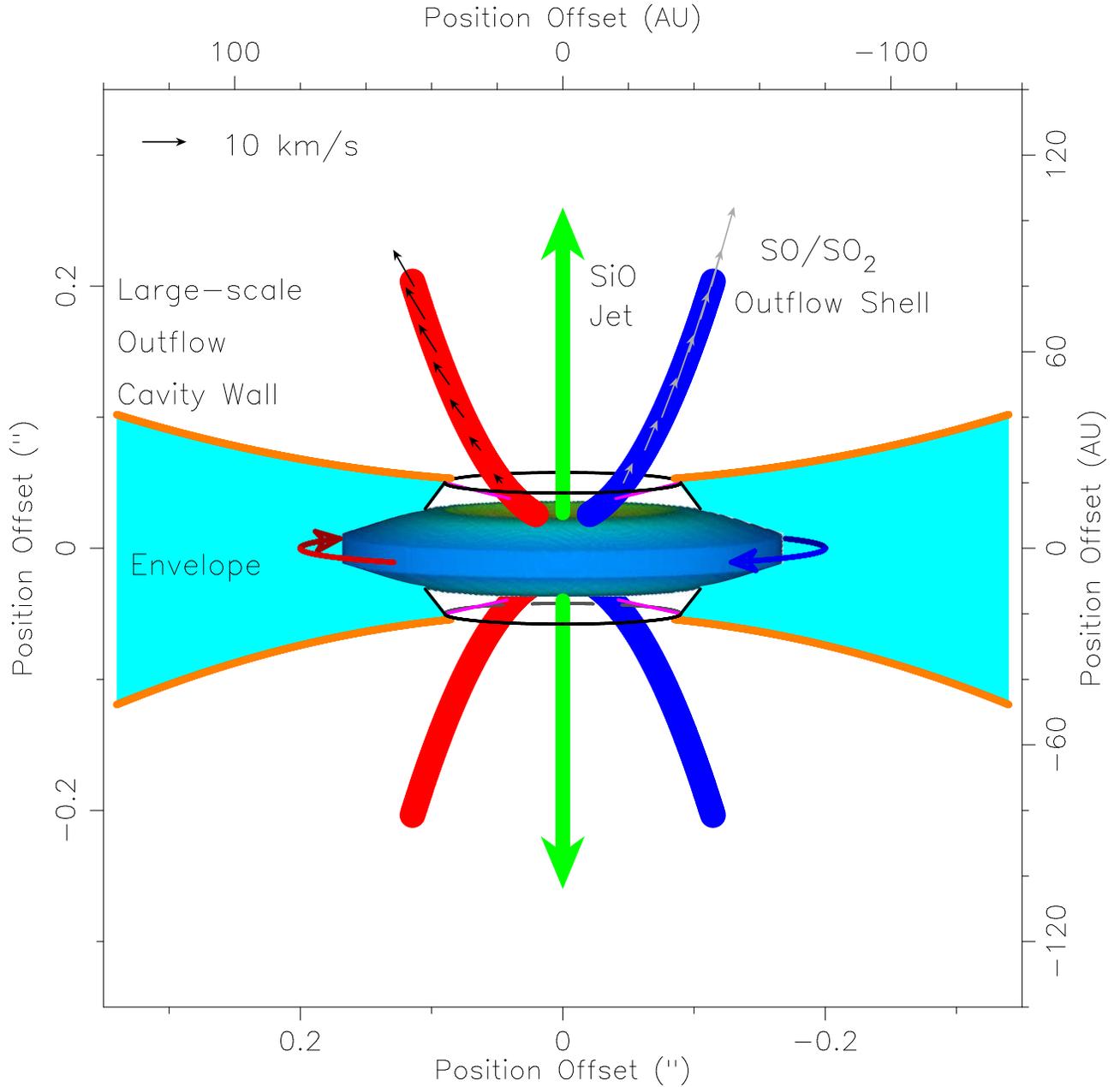

\centering
\putfig{0.9}{270}{f10.eps} 
\figcaption[]
{Schematic diagram showing the wide-opening
SO/\SOO{} outflow shell in connection to the SiO jet, and other components 
(e.g., infalling-rotating envelope, dusty disk, disk atmosphere, and
large-scale outflow cavity walls) presented
in Figure 9b of \citet{Lee2017com}.
Black arrows show the velocity vectors in the radially expanding shell model,
while gray arrows show the velocity vectors in the material moving along the shell model.
\label{fig:model}}
\end{figure}

Recently, a compact outflow has been detected in HH 212 at $\sim$ 52 au
(\arcsa{0}{13}) resolution, extending out from the disk \citep{Tabone2017}. 
It is rotating and thus may trace a disk wind.  Now at $\sim$ 16 au
(\arcsa{0}{04}) resolution, the compact outflow is found to consist of two
distinct components: a wide-angle outflow shell and a collimated jet. 
The collimated jet is aligned with the SiO jet and thus likely traces
the shock interactions along the jet axis.  Interestingly, the outflow shell
lies within the large-scale molecular outflow detected in C$^{34}$S
\citep{Codella2014} and \HCOP{} \citep{Lee2017com}, and is thus not the
rotating envelope material pushed away by the underlying wind or jet (see
Figure \ref{fig:model} for various components).  Moreover, it is rotating
and seen extending out from the inner disk at a radius of $<$ 20 au within
the centrifugal barrier \citep{Lee2017com}, and thus may trace a disk wind
coming out from there.  The specific angular momentum is roughly constant at
$\sim$ 40 au \vkm{} within $\sim$ 50 au of the central source.  This
specific angular momentum corresponds to that at $\sim$ 7.2 au in the disk,
where the rotation velocity is $\sim$ 5.5 \vkm{}, assuming a central mass of
0.25 \solarmass{} \citep{Lee2017com}.  It decreases to $\sim$ 25 au \vkm{}
at a distance of $\sim$ 80 au.  This decrease in specific angular momentum
could be inconsistent with that expected for a disk-wind, however, further
observations are needed to confirm this.  On the other hand, it is also
possible that the SO/\SOO{} outflow shell is the disk/atmosphere material
pushed up and out by a wind launched at the innermost disk, e.g., by a
wide-angle radial wind component surrounding the fast jet as in the X-wind
model.

\begin{figure} [!hbp]
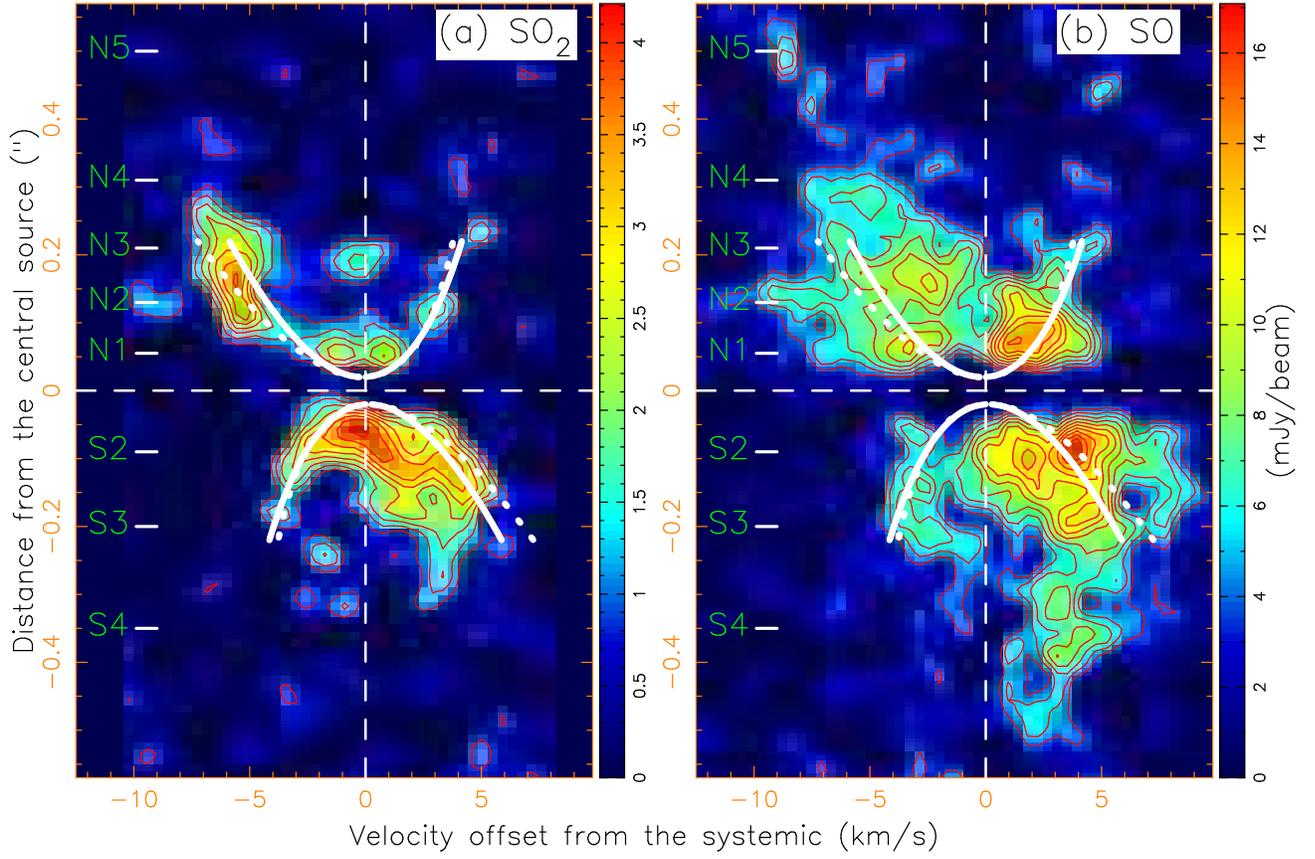

\centering
\putfig{0.65}{270}{f11.eps} 
\figcaption[]
{Same as Figure \ref{fig:pvjet}, but with two model PV structures added.
The solid curves are for the radial
expanding shell model with $v_0 \sim 42$ \vkm{} arcsec$^{-1}$ and the dotted
curves are for the material moving along the shell model
with $v_0 \sim 45$ \vkm{} arcsec$^{-1}$.
\label{fig:pvjetfit}}
\end{figure}

The shell kinematics could be used to differentiate the above two origins of
the outflow shell.  If the shell traces the disk wind itself, the velocity
vectors are expected to be parallel to the shell structure.  If the shell
traces the disk or atmosphere material pushed away by a wide-angle radial
wind, then the velocity vectors in the shell are expected to be radially
directed.  For simplicity, the shell here can be assumed to roughly
have a parabolic structure with $z = C R^2+$\arcsa{0}{02} in cylindrical
coordinate system and $C \sim$ 14 arcsec$^{-1}$ (as indicated by the white
curves in Figure \ref{fig:mom1}). If the shell material is moving along
the shell, then we have $v_R = v_0 R$ and $v_z = 2 v_0 z$ in cylindrical
coordinate system.  If the shell material is radially expanding, then we
have $v_R = v_0 R$ and $v_z = v_0 z$ \citep{Lee2000}.  Figure
\ref{fig:pvjetfit} shows the model PV structures of these two models at an
inclination angle of $\sim$ 4\degree{}, in comparison to the observed PV
structures.  The solid curves are from the radially expanding shell model
with $v_0 \sim 42$ \vkm{} arcsec$^{-1}$ and the dotted curves are from the
material moving along the shell model with $v_0 \sim 45$ \vkm{}
arcsec$^{-1}$.  The velocity vectors and magnitudes in these two models are
shown in Figure \ref{fig:model}, with the black arrows for the former and
gray arrows for the latter.  As can be seen, both model PV structures can
match the observed PV structure reasonably well.  Therefore, unfortunately,
we can not determine which model is better based on the kinematics here in
HH 212.  Having said that, it is unclear whether a disk wind can have an
expansion speed that increases with the distance from the disk and at the
same time a specific angular momentum that decreases with distance.  For a
shell accelerated by an underlying wide-angle radial wind, it is reasonable
to have an expansion velocity increasing with the distance
\citep{Shu1991,Lee2000}, because the disk atmosphere has a density
decreasing with height.  Whether the specific angular momentum can naturally
decrease with distance in this scenario is uncertain.  In addition, the
shell in the north is seen connecting to knot N4 further north, likely also
affected by it (see Figure \ref{fig:so}).





Another question we can ask is that: does the wide-angle wind component of
an X-wind, if exists, have enough momentum to drive the outflow shell? 
The mass of the shell can be roughly estimated from the mean H$_2$ column
density derived from the SO emission in Section \ref{sec:colden}. Thus, the
total mass of the shell within $\sim$ \arcsa{0}{2} of the central source in
the north and south is $M_s = 1.4 \mHt \NHt A \sim$ $(0.4-4)\times10^{-4}$
\solarmass{}, where $A$ is the area of the shell.  Assuming a mean expansion
velocity of $\sim$ 5 \vkm{} in the shell, the momentum is then
$(0.2-2)\times10^{-3}$ \solarmass{} \vkm{}.  In the X-wind model, the
wide-angle wind component could have a mass-loss rate and a velocity similar
to those of the jet component \citep{Shu1995}.  The jet component was found
to have a mass-loss rate of $\sim$ $10^{-6}$ \solarmass yr$^{-1}$ and a
velocity of $\sim$ 100 \vkm{} \citep{Lee2015}.  According to our simple
radially expanding model for the shell, the shell has a dynamical age of
$1/v_0$ or $\sim$ 45 yrs, indicating that the shell could have been driven
for that long by an underlying wind. Thus, the wide-angle wind component
can in principle provide an accumulated momentum of $\sim$
4.5$\times10^{-3}$ \solarmass{} \vkm{}, enough to drive the shell.


One more question to ask is that: why is there only one shell?  In the case
of a radial expanding shell, the shell has a dynamical age much younger than
the age of HH 212.  Since the jet is episodic, the inner wind that drives
the shell should also be episodic.   Since the semi-periodic spacing
between two bow shocks in the jet is $\sim$ \arcs{3} \citep{Lee2007}, the
period for the major ejection of the jet/wind is $\sim$ 50 yrs and thus
slightly longer than the dynamical age of the shell. Therefore, it is
possible that the shell is produced recently and the earlier shells are lost
in the outflow cavity probably because the density becomes too low to excite
the SO and \SOO{} lines.  Or as suggested in \citet{Tabone2017}, the shell
traces an onionlike shell in the disk wind coming out from the disk around
the SiO jet (see their Figure 3), but the part further away from the central
source is lost in the cavity.

Independent of the exact mechanism for producing the outflow shell, if it is
really rotating with a specific angular momentum of $\sim$ 40 au \vkm, it is
likely coming from the disk at a relatively large radius of at least several
au.   If the outflow is indeed a MHD disk wind from the disk, then
using Eq.  4 in \citet{Anderson2003}, and an outflow velocity of 5-10
\vkm{}, the launching radius is estimated to be 2-3 au. The specific
angular momentum is much larger than the mean value for the SiO jet ($\sim$
10 au \vkm), indicating that the outflow shell is not simply the jet
material that is swept into a shell.  One may argue that the shell material
may come from the infalling protostellar envelope that extends all the way
to less than several au above and below the disk.  However, the large-scale
outflow cavity wall, which presumably marks the inner boundary of the
envelope, is located well outside the outflow shell (as indicated in the
cartoon of Figure \ref{fig:model}).  Furthermore, in the case of much more
evolved T Tauri star HH 30, there is little evidence for a large-scale
envelope, and yet a rotating, expanding shell is still observed surrounding
the jet (Louvet et al.  in prep), indicating that an envelope is not needed
for producing such a structure.  This leaves the material coming directly
from the disk as the most likely material for the outflow shell.

The key question is whether that disk material in the outflow shell is
launched from the several au radius by some mechanism internal to the local
disk, such as through magnetocentrifugal acceleration or magnetic pressure
gradient, or by some external agents, such as interaction with a wide-angle
wind surrounding the SiO jet.  Indeed, the two mechanisms are not mutually
exclusive: outflow launched from several au radius is expected to interact
with the wide-angle wind interior to it, if such a wind exists.  In any
case, the rotating outflow shell carries away angular momentum from the
disk, most likely from the several au radius, which is in contrast with the
rotating SiO jet driven from $\sim$ 0.05 au radius.





\subsection{Formation of SO and SO$_2$}



Previous observations of HH 212 at an angular resolution of $\sim$
\arcsa{0}{5} have shown that the SO and \SOO{} abundances in the outflow are
very high, up to $\sim$ 10$^{-6}$ \citep{Podio2015}.  Now at higher
resolution, the outflow is resolved into a small-scale wide-opening outflow
shell and a collimated jet.  The high SO and \SOO{} abundances in the
collimated jet could be due to violent (internal) shocks in the jet.  The
origin of high abundances in the wide-angle outflow shell is uncertain.  If
the shell traces a disk wind launched from the disk, then the high
abundances could be due to the thermal sublimation of icy mantles near the
source and the gradual heating by ambipolar diffusion during MHD
acceleration \citep{Panoglou2012}.  If the shell is the disk/atmosphere
material pushed away by an inner wind, then the high abundances could be due
to shock interactions with the inner wind.

As for the disk, \citet{Podio2015} have found a high SO abundance of
10$^{-8}$--10$^{-7}$.  Our observations also show high SO and \SOO{} column
densities (see Section \ref{sec:colden}) and thus high SO and \SOO{}
abundances there in the disk atmosphere.  These high abundances in the disk
atmosphere could be due to the accretion shocks already observed in HH 212
by \citet{Lee2017jet}, radiation heating by the central protostar, or shock
interactions with an inner wind.

\section{Conclusions}

We have mapped the central 400 au region in the HH 212 protostellar system
in SO and \SOO{}, with the Atacama Large Millimeter/submillimeter Array at
$\sim$ 16 au resolution.  We have resolved the compact outflow near the
central source into a wide-opening rotating outflow shell with a width of
$\sim$ 100 au and a collimated jet.  The collimated jet is aligned with the
SiO jet and thus likely traces the shock interactions along the jet axis. 
The wide-opening outflow shell is seen around the base of the collimated SiO
jet, extending out from the inner disk.  It is not only expanding away from
the center, but also rotating around the jet axis.  The specific angular
momentum of the outflow shell is $\sim$ 40 au \vkm{}.  Simple modeling
suggests that this rotating outflow shell can trace either a disk wind or
disk material pushed away by an unseen inner wind.  We also resolve the disk
atmosphere traced by the two S-bearing molecules, confirming the Keplerian
rotation there.

\acknowledgements

We thank the anonymous referee for useful comments.
This paper makes use of the following ALMA data: ADS/JAO.ALMA\#
2015.1.00024.S.  ALMA is a partnership of ESO (representing its member
states), NSF (USA) and NINS (Japan), together with NRC (Canada), MoST and
ASIAA (Taiwan), and KASI (Republic of Korea), in cooperation with the
Republic of Chile.  The Joint ALMA Observatory is operated by ESO, AUI/NRAO
and NAOJ.  C.-F.L.  acknowledges grants from the Ministry of Science and
Technology of Taiwan (MoST 104-2119-M-001-015-MY3) and the Academia Sinica
(Career Development Award).  ZYL is supported in part by NASA NNX14AB38G and
NSF AST-1313083 and 1716259.




\def\nat{Nature}

\begin{table}
\small
\centering
\caption{Observation Logs}
\label{tab:obs}
\begin{tabular}{llllll}
\hline
Cycle & Date & Array &Number of &Time on target & Baselines  \\
      & (YYYY-MM-DD) & Configuration & Antennas&(minutes) & (meter) \\
\hline\hline
3 & 2015-11-05 & C36-7/8 & 45 & 44  & 78$-$16196 \\ 
3 & 2015-12-03 & C36-7/8 & 32 & 44  & 17$-$6344  \\ 
\hline
\end{tabular}
\end{table}

\begin{table}
\small
\centering
\caption{Correlator Setup for Cycle 3 Project}
\label{tab:corr3}
\begin{tabular}{llllll}
\hline
Spectral  & Line or   & Number of & Central Frequency & Bandwidth & Channel Width\\
Window & Continuum & Channels  & (GHz)             & (MHz)     & (kHz) \\
\hline\hline
0 & SO \SOt      & 960   & 346.528 & 234.375  & 244.140  \\
1 & CO $J=3-2$      & 960   & 345.796 & 234.375  & 244.140  \\
2 & H$^{13}$CO$^+$ $J=4-3$      & 960   & 346.998 & 234.375  & 244.140  \\
3 & SiO $J=8-7$     & 960   & 347.330 & 234.375  & 244.140  \\
4 & HCO$^+$ $J=4-3$ & 1920  & 356.735 & 468.750  & 244.140  \\
5 & Continuum     & 1920  & 357.994 &1875.000  & 976.562  \\
\hline
\end{tabular}
\end{table}

\begin{table}
\small
\centering
\caption{Calibrators and Their Flux Densities}
\label{tab:calib}
\begin{tabular}{llll}
\hline
Date & Bandpass Calibrator &Flux  Calibrator & Phase Calibrator \\
(YYYY-MM-DD) & (Quasar, Flux Density) & (Quasar, Flux Density) & (Quasar, Flux Density) \\
\hline\hline
2015-11-05 & J0423-0120, 0.55 Jy & J0423-0120, 0.55 Jy & J0541-0211, 0.22 Jy \\  
2015-12-03 & J0510+1800, 4.07 Jy & J0423-0120, 0.67 Jy & J0541-0211, 0.23 Jy \\ 
\hline
\end{tabular}
\end{table}

%
\begin{deluxetable}{lcccccc}
\tablecolumns{8}
\tabletypesize{\normalsize}
\tablecaption{Line Properties from Splatalogue
 \label{tab:lines}}
\tablewidth{0pt}
\tablehead{
\colhead{Molecule} & \colhead{Frequency} & Transition & 
$S_{ij}\mu^2$ &     log$_{10}$($A_{ij}$) &     $E_{up}$ & Linelist \\
               & \colhead{(GHz)}   & QNs  & (D$^2$)            &   (s$^{-1}$) & (K) & 
}  
\startdata
 SO   & 346.52848 & 	9(8)- 8(7) &	21.52703 &	-3.26062 &	78.77510 & 	JPL\\
 \SOTwo  & 356.75519 & 10(4, 6)-10(3, 7) & 13.03465 &	-3.48408 & 89.83365  &	JPL \\   
 \SOTwo  & 357.16536 & 13(4,10)-13(3,11) & 17.87923 &	-3.45447 & 122.96459 & 	JPL \\
 \SOTwo  & 357.24119 & 15(4,12)-15(3,13) & 21.14165 &	-3.44140 & 149.68170 & 	JPL \\
 \SOTwo  & 357.38757 & 11(4, 8)-11(3, 9) & 14.64155 &	-3.47079 & 99.95222  &	JPL \\
 \SOTwo  & 357.58145 &  8(4, 4)- 8(3, 5) & 9.76046  &	-3.51492 & 72.36335  &	JPL \\
 \SOTwo  & 357.67178 &  9(4, 6)- 9(3, 7) & 11.40052 & 	-3.49544 & 80.63686  &	JPL \\
 \SOTwo  & 357.89244 &  7(4, 4)- 7(3, 5) & 8.08092  &	-3.54144 & 65.01106  &	JPL \\
 \SOTwo  & 357.92596 &  6(4, 2)- 6(3, 3) & 6.34115  &	-3.58446 & 58.57965  &	JPL \\
 \SOTwo  & 357.96289 & 17(4,14)-17(3,15) & 24.42371 &	-3.42881 & 180.11104 & 	JPL \\
 \SOTwo  & 358.01309 &  5(4, 2)- 5(3, 3) & 4.49029  &	-3.66149 & 53.06831  &	JPL \\
 \SOTwo  & 358.03808 &  4(4, 0)- 4(3, 1) & 2.44211  &	-3.83876 & 48.47653  &	JPL \\
 \SOTwo  & 358.21564 & 20(0,20)-19(1,19) & 44.65235 &	-3.23457 & 185.32964 &	JPL \\
\enddata
\end{deluxetable}
\clearpage

\end{document}